\documentclass[%
onecolumn, superscriptaddress,
showpacs,preprintnumbers,
nofootinbib,
 amsmath,amssymb,
 aps,
pra,
]{revtex4}
\usepackage[all]{xy}
\usepackage{mathrsfs}
\usepackage{graphicx}% Include figure files
\usepackage{epstopdf}
\usepackage{dcolumn}% Align table columns on decimal point
\usepackage{bm}% bold math
\usepackage{color} % change color of text
\usepackage{CJK}
\def\la{{\langle}}
\def\ra{{\rangle}}

\newcommand{\beq}{\begin{equation}}
\newcommand{\eeq}{\end{equation}}
\newcommand{\beqa}{\begin{eqnarray}}
\newcommand{\eeqa}{\end{eqnarray}}

\begin{document}
\title{Shortcuts to adiabaticity for an ion in a rotating radially-tight trap}
\author{M. Palmero}
\email{mikel.palmero@ehu.eus}
\affiliation{Departamento de Qu\'{\i}mica F\'{\i}sica, UPV/EHU, Apdo.
644, 48080 Bilbao, Spain}
\author{Shuo Wang}
\affiliation{Electrical and Systems Engineering, Washington University in St. Louis, St. Louis, Missouri 63130, USA}
%
%\author{Kazutaka Takahashi}
%\affiliation{Department of Physics, Tokyo Institute of Technology, Tokyo 152-8551, Japan}
%
\author{D. Gu\' ery-Odelin}
\affiliation{Laboratoire de Collisions Agr\' egats R\' eactivit\' e, CNRS UMR 5589, IRSAMC, Universit\' e de Toulouse (UPS), 118 Route de Narbonne, 31062 Toulouse CEDEX 4, France}
\author{Jr-Shin Li}
\affiliation{Electrical and Systems Engineering, Washington University in St. Louis, St. Louis, Missouri 63130, USA}
\author{J. G. Muga}
\affiliation{Departamento de Qu\'{\i}mica F\'{\i}sica, UPV/EHU, Apdo.
644, 48080 Bilbao, Spain}
\affiliation{Department of Physics, Shanghai University, 200444
Shanghai, People's Republic of China}
\begin{abstract}
We engineer the fast rotation of an effectively one-dimensional ion trap  
for a predetermined rotation angle and time, avoiding the final excitation of the trapped ion.  
Different schemes are proposed with different speed limits that depend on the control capabilities.  
We also make use of trap rotations to create squeezed states without manipulating
the trap frequencies. 
\end{abstract}
%\pacs{37.10.Gh, 02.30.Yy}
%\pacs{37.10.Gh, %Atom traps and guides
%02.30.Yy, % Control theory
%03.65.Ca, %Formalism
%03.65.Nk % Scattering theory
%37.10.Ty Ion trapping
%}
\maketitle
%
%
%
%
%\tableofcontents
%
\section{Introduction}
%
%
% % % % % % % % % % % % % % % % % % % % % % % % % % % % % % % % % % % % % % % % %
% % % % % % % % % % % % % % % % % % % % % % % % % % % % % % % % % % % % % % % %
\begin{figure}[b]
\begin{center}
\includegraphics[width=7cm]{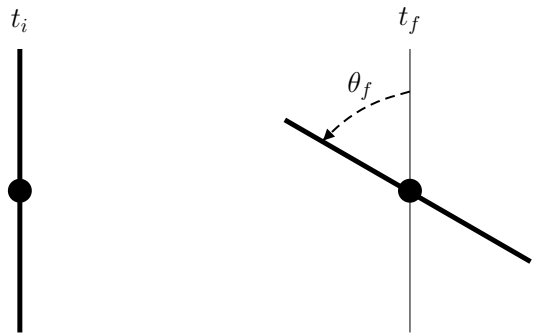}
\caption{\label{scheme}(Color online)
Schematic representation of the rotation process. The ion is confined along a line (where it 
is subjected to an effective   one-dimensional -longitudinal- potential), which is rotated by an angle $\theta$ up to $\theta_f$ in a time $t_f$, 
so that the final state is not excited.}
\end{center}
\end{figure}
% % % % % % %º º%º ºº1\\\\\\% \\\% % % \% % % % % % % % % % % % % % % % % % % % % % % % % % % % % % %
% % % % % % % % % % % % % % % % % % % % % % % % % % % % % % % % % % % % % % % % % % %
%
A major challenge in modern atomic physics is to develop a scalable architecture for quantum information
processing. A proposed scheme to achieve scalability is based on shuttling individual ions or small groups of ions 
among processing or storing sites  \cite{Wineland2002,Rowe,Wineland2006,Home,Roos,Monroe}.  
Apart from shuttling \cite{Torrontegui,Palmerotrans,Palmerotrans2,Lu1,Lu2},  other manipulations of the ion motion would be needed, 
such as expansions or compressions of the trap \cite{Chen,Palmeroexp}, separating or merging ion chains \cite{Lau2012,Palmerosep,Kaufman}, and rotations \cite{Splatt}. 
All these basic dynamical operations should fulfill two seemingly contradictory requirements: 
they should be fast, but free from final motional excitation. Shortcuts to adiabaticity for ``fast and safe driving'' 
have been proposed for several of these elementary operations \cite{Torrontegui,Palmerotrans,Palmerotrans2,Lu1,Lu2,Chen,Palmeroexp,Palmerosep} and have also been implemented experimentally \cite{Bowler,Schmidt}.

In this paper we study rotations of a single ion as depicted in Fig. \ref{scheme}. Our aim is to  inverse engineer 
the time-dependence of the control parameter(s) to implement a fast  process, free from final excitations. 
We assume for simplicity that  the ion is trapped in a linear, harmonic trap, tightly confined in a radial 
direction so that it  moves effectively along a one-dimensional axial direction, hereafter ``the line''. 
The trapping line is
set horizontally  and is rotated in a time $t_f$ up to an established final angle ($\theta_f= \pi/2$ in all examples) with respect to a
vertical axis that crosses the center of the trap.           
Such an operation would be useful to drive atoms through corners and junctions in  a scalable quantum processor \cite{Hensinger, Amini}. 
It is also a first  step towards the more complicated problem of  rotating an ion chain \cite{Hensinger,swapNIST,Splatt}, 
which would facilitate scalability in linear
segmented traps, and be useful to rearrange the ions, e.g. to locate a cooling ion at the right position in the chain \cite{Splatt}.

%In this section vi\de the Hamiltonian and the notation that describes the physical setting. 
%As the ion's motion is constrained to a rotating, one-dimensional, horizontal  line, 
We shall first find the classical Hamiltonian.   
Let $s$ denote a point on the line. $s$ 
may take positive and negative values.  
A time dependent trajectory $s(t)$ has Cartesian, laboratory frame components $x=x(s,t)$,
$y=y(s,t)$,    
\beq
x=s\cos (\theta),
\; 
y=s\sin (\theta),
\eeq
where $\theta=\theta (t)$ is the rotation angle.  
The kinetic energy is $K=\frac{1}{2}m(\dot{x}^2+\dot{y}^2)$, where $m$ is the ion mass,  and the potential energy 
is assumed by now to be harmonic, $\frac{1}{2}m\omega_0^2 s^2$ (this will be relaxed below and in Sec.  \ref{ctf}), 
where $\omega_0$ is the angular frequency 
of the external confining trap in the (longitudinal) direction of the line.
This gives the Lagrangian 
\beqa
L &=& \frac{1}{2} m\dot{s}^2-\frac{1}{2}m \omega^2 s^2,
\\
\omega^2 &=& \omega_0^2 -\dot{\theta}^2.
\label{omeef}
\eeqa
Note that the angular velocity of the rotation $\dot{\theta}$  must be real but could be negative, whereas $\omega^2$ 
may be positive or negative,  making $\omega$ purely imaginary in the later case.    
Unless stated otherwise, the following physically motivated boundary conditions are also assumed:  
the initial and final trap should be at rest, and we also impose continuity of  
the angular velocity, 
\beqa
\theta(0)&=&0,\quad \theta(t_f)=\theta_f,
\label{cond1}\\
\dot{\theta}(0)&=&\dot{\theta}(t_f)=0,
\label{cond2}\\
\omega(0)&=&\omega(t_f)=\omega_0,
\label{cond3}
\eeqa 
where the last line follows  from the second one using Eq. (\ref{omeef}).   
By a Legendre transformation we finally get the Hamiltonian\footnote{This is easily generalized for a potential $U(s)$, not necessarily harmonic,  as $H= \frac{1}{2} m\dot{s}^2+U(s)-\frac{1}{2}m\dot{\theta}^2 s^2$} 
\beq
H= \dot{s}\frac{\partial L}{\partial\dot{s}}-L=\frac{1}{2} m\dot{s}^2+\frac{1}{2}m\omega^2 s^2.  
\eeq
%  
%which is we regard from now on as our Hamiltonian determining the dynamics of the quantum particle. 
At this point, we quantize this Hamiltonian by substituting $m\dot{s}$ by the momentum operator 
$p$ and by considering $s$ as the position operator, which becomes a $c$-number in coordinate representation, 
\beq\label{Ham}
H= \frac{1}{2m} {p}^2+\frac{1}{2}m\omega^2 s^2.  
\eeq 
%. 
We will from now on work with this  quantum Hamiltonian (possibly with a more general potential) 
and corresponding quantum states.  
It represents formally a harmonic oscillator with time-dependent frequency, but there are 
significant differences with an actual harmonic oscillator 
when the inverse engineering of $\omega (t)$ is considered.  For an actual harmonic oscillator   
a fast and safe expansion or compression in a time $t_f$ should take the system from
an initial value to a final value of $\omega$ without final excitation, in principle  without further
conditions.  
By contrast, in the rotation process, according to Eq. (\ref{cond3}), the initial and final effective 
frequencies are the same, but  the conditions in Eqs. (\ref{cond1}) and (\ref{cond2}) must be
satisfied.  This implies an integral constraint on $\omega$, 
%From the definition in Eq. (\ref{Omega})
%We get,  by integration, 
%
\beq
\theta(t_f)=\int_0^{t_f} \dot\theta dt' = \int_0^{t_f} [\omega_0^2-\omega^2]^{1/2}  dt',
\label{omecon}
\eeq
where the square root branch should be chosen to satisfy continuity.  
%and transient changes in $\omega (t)$.
One further difference is that in a physical expansion/compression $\omega (t)$ is controlled directly
whereas in the rotation there are several options. If $\omega_0$ is constant, only  
$\dot{\theta} (t)$ is controlled, so that $\omega (t)$ is an `effective' frequency.
In general both $\omega_0$ and $\dot{\theta}$ could be controlled as time-dependent functions, see the next section.       
%To make $\omega(t_f)=0$, see Eq. (\ref{Omega}),
%
%\beq
%\int_0^{t_f} \Omega dt'=0.
%\label{Omecon1}
%\eeq
% 
%From the definition in Eq. (\ref{Omega})
%We get,  by integration, 
%%
%\beq
%\theta(t_f)=\int_0^{t_f} \omega(t')  dt'.
%\label{omecon}
%\eeq
%%
As for the final excitation,  the expression for the energy 
of a state that begins in the $n$-th eigenstate of the trap at rest 
can be found making use of the Lewis-Riesenfeld invariants  \cite{LR, Chen}, see the corresponding time-dependent wave function in the Appendix, 
\beq
\label{energy}
\la H(t)\ra_n=\frac{(2n+1)\hbar}{4 \omega (0)}\left(\dot{b}^2+\omega ^2(t)b^2+\frac{\omega (0)^2}{b^2}\right).
\eeq
Here $b$ is a scaling factor, proportional to the width of the invariant eigenstates,
that satisfies the Ermakov equation
\beq 
\label{differential}
\ddot{b}+\omega ^2(t) b=\frac{\omega^2(0)}{b^3}.
\eeq
To avoid any final  excitation, it is required that  
\beq
b(t_f)=1,\quad \dot{b}(t_f)=0
\label{finalcb}
\eeq
for the initial conditions $b(0)=1,\;\dot{b}(0)=0$. 
%When inverse-engineering the rotation, $b$ may be designed subjected to these 
%conditions and such that the integral condition for $\omega$ is satisfied. 
The boundary conditions for $b$ and Eqs. (\ref{cond1},\ref{cond2},\ref{cond3}) imply  that $H(0)=H(t_f)$ commutes with the corresponding Lewis-Riesenfeld invariant \cite{LR}, so that the $n$-th initial eigenstate  is dynamically mapped 
onto itself (but rotated) at time $t_f$.   In Eqs. \eqref{energy} and \eqref{differential} both the excitation energy and the wave packet width are mass independent, so that inverse-engineered rotation protocols will be independent of the species. 
%???
%
%\beq
%|\psi _n \rangle=\frac{1}{b^{1/2}} e^{\frac{i}{\hbar} \frac{\dot{b}q^2}{2 b}} \Phi_n \left(\frac{q}{b}\right),
%\eeq
%
%where the $\Phi_n$ are the eigenstates of a static harmonic oscillator.
%Of course for arbitrary rotations satisfying Eqs. (\ref{cond1}-\ref{cond3}) these excitation-free conditions 
%for $b$ in Eq. (\ref{finalcb}) are not satisfied. 
%
In the following sections we shall analyze different methods to perform the  rotation 
without final excitation.
%

%
% % % % % % % % % % % % % % % % % % % % % % % % % % % % % % % % % % % % % % % % %
% % % % % % % % % % % % % % % % % % % % % % % % % % % % % % % % % % % % % % % % % % %
%
%
%
%
%
% 
\section{Control of trap frequency and angular velocity\label{ctf}}
If both the trap angular frequency $\omega_0$ and the angular velocity $\dot{\theta}$ 
are controllable functions of time,
a simple family of solutions to the inverse problem is found by setting a $\dot{\theta} (t)$ that satisfies Eqs. (\ref{cond1}) and (\ref{cond2}), and compensating 
the time dependence of $\dot{\theta}^2$ with a corresponding change in
$\omega_0^2(t)$, 
so that $\omega^2(t)=\omega^2(0)$ remains constant during the whole process. 
From the point of view of the effective 
harmonic oscillator `nothing happens' throughout the rotation, so that the effective state
remains unexcited at all times.
% A large family of $\omega(t)$ functions may be found to satisfy   Eqs. (\ref{cond1}) and (\ref{cond2}).     
%
% % % % % % % % % % % % % % % % % % % % % % % % % % % % % % % % % % % % % % % % %

% % % % % % % % % % % % % % % % % % % % % % % % % % % % % % % % % % % % % % % % % % %
%    

We may apply the Lewis-Leach theory of quadratic in momentum invariants \cite{LL,DL} to extend the above results to arbitrary 
potentials\footnote{The theory was first formulated for classical systems in \cite{LL} but is applicable to
quantum systems as well \cite{DL}. Incidentally this means that the rotation protocols designed in this paper -in this and the following sections- are valid for classical particles as well. The difference appears only when considering which states are valid or  not for classical and quantum particles, e.g., when using phase-space formulations of quantum states and classical ensembles.}.
The family of Hamiltonians 
\beq\label{hami}
H=\frac{p^2}{2m}+\frac{1}{2} m \Omega^2 s^2+ \frac{1}{b^2} U\left(\frac{s}{b}\right), 
\eeq
where $U$ is an arbitrary function, and $\Omega$ depends on time, 
has the 
invariant
\beq
I=\frac{\pi^2}{2m}+\frac{1}{2} m \Omega_0^2 s^2+ U\left(\frac{s}{b}\right),  
\eeq
where $\pi=bp-m\dot{b}s$, and $\Omega_0$ is a constant, provided the Ermakov equation 
\beq\label{Erm}
\ddot{b}+\Omega^2 b=\frac{\Omega_0^2}{b^3}
\eeq
is satisfied. 
Consider the simple case $\Omega_0=0$, 
i.e., from  Eq. (\ref{Erm}), 
\beq
\Omega^2(t)=-\frac{\ddot{b}}{b}.
\eeq
% 
%and   
%
%\beqa
%b(0)=1,\, b(t_f)=1,
%\\
%\dot{b}(0)=\dot{b}(t_f)=0,
%\eeqa
%
If we set $b(t)=1$ as a constant for all times, it follows  that 
$\Omega(t)=0$.
However, as we saw in the previous section, the rotation of a line with the potential $U(s)$ produces in the line frame a centrifugal term 
$-\dot{\theta}^2s^2m/2$. To cancel the total harmonic term, we have to add to the trap potential a 
compensating harmonic term, $\omega_c^2s^2m/2$, 
such that $\omega_c^2=\dot\theta^2$.  In other words, $\Omega^2=\omega_c^2-\dot{\theta}^2=0$. 
The resulting Hamiltonian and invariant (in this case they are equal) are simply 
\beq
H=I=\frac{p^2}{2m}+U(s),
\eeq
i.e., time independent. No excitation occurs at any time in spite of the fact that a rotation is taking place.  

For some applications it may be interesting to consider in Eq. (\ref{hami}) the more general case in which $b$ depends on time
(for example to achieve a squeezed state), and $\omega^2=\omega_c^2-\dot{\theta}^2$, corresponding to an auxiliary harmonic term
and the centrifugal term.  
The inverse engineering in this case proceeds by designing $\theta(t)$, so that $\dot{\theta}(0)=\dot\theta(t_f)=0$,
and then $b(t)$ 
obeying the boundary conditions
\beqa
b(0)&=&b(t_f)=1,
\\
\dot{b}(0)&=&\dot{b}(t_f)=0,
\\
\ddot{b}(0)&=&\ddot{b}(t_f)=0,
\eeqa 
(or more generally $b(t_f)=\gamma$) that guarantee the commutation between invariant and Hamiltonian at boundary times. 
Once $\theta$ and $b$ are set we design the auxiliary harmonic term considering, as before, $\Omega_0=0$ in Eq. (\ref{Erm}),      
\beq
\omega_c^2=\Omega^2+\dot{\theta}^2=-\frac{\ddot{b}}{b}+\dot{\theta}^2.
\eeq
The auxiliary harmonic term vanishes at both boundary times according to the boundary conditions imposed on $\ddot{b}$
and $\dot{\theta}$. In fact $\Omega^2$ vanishes as well at the boundary times so that 
before and after the rotation the atom is confined only in the potential $U(s)$. 
%In addition to the harmonic term $m\Omega^2 s^2/2$ the potential on the line must 
%depend on time as $\frac{1}{b^2}U(\frac{s}{b})$,   so that the effective (rotating frame) 
%Hamiltonian takes the form of Eq. (\ref{hami}).   
%, which might be interesting rather than having it confined in the harmonic potential.
%In addition to the need of producing the harmonic term, the trap potential must 
%scale as $b^{-2}U(x/b)$ .  
This type of protocols, where both the rotation speed and the potential have to be 
controlled (the latter in space and time) may be quite demanding experimentally. In the rest of the paper we 
shall assume  the simpler scenario in which only the rotation speed $\dot{\theta}$ is controlled, and the 
trap potential is purely harmonic with constant angular frequency $\omega_0$.   
%
%\beqa
%b(0)=1,\, b(t_f)=1,
%\\
%\dot{b}(0)=\dot{b}(t_f)=0,
%\eeqa     
%
% 
%
%\section{Optimal Control}
%
%
%
%
%
%
%
\section{Bang-bang}
It is possible to perform rotations without final excitation 
satisfying Eqs. (\ref{cond1}) and (\ref{cond2})
keeping  
$\dot{\theta}$  constant or piecewise constant. Here we consider the simplest one-step case,  
\beq
\dot{\theta}(t)=\left\{
\begin{array}{ll}
0,&t\le 0
\\
c,& 0<t \le t_f
\\
0,& t\ge t_f
\end{array}
\right..
\eeq
Note that Eqs. (\ref{cond2}) and (\ref{cond3}) are only satisfied now as one-sided limits. 
A bang-bang approach may admitedly be difficult to implement because of the sharp changes involved, 
but it sets a useful, simple reference for orders of magnitude estimates of rotation speeds 
which may be compared to smoother approaches 
that will be presented later.      
%The discontinuity of $\dot{\theta}$ is however a relatively mild one, 
%in the sense that the basic equations ... still hold ???. 
Integrating $\dot\theta$ we find
\beq
\theta_f=c t_f.
\label{thetaf}
\eeq
For a constant $\dot{\theta}=c$, $\omega$ remains constant from $t=0$ to $t=t_f$, and equal to 
$\omega_1=(\omega_0^2-c^2)^{1/2}$, whereas $\omega=\omega_0$ in the initial and final
time regions. 
For this configuration, and $0<t<t_f$, 
\beqa
b(t)&=&\sqrt{\frac{\omega_0^2-\omega_1^2}{\omega_1^2}\sin^2(\omega_1 t)+1},
\label{bbb}
\\
\dot{b}(t)&=&\frac{\sin(\omega_1 t)\cos(\omega_1 t)(\omega_0^2-\omega_1^2)}{\omega_1b(t)},
\eeqa
to satisfy the boundary conditions $b(0)=1$, $\dot{b}(0)=0$. 
The shortest final time to satisfy the conditions (\ref{finalcb}) at $t_f$ is $\pi/\omega_1$. 
From Eq. (\ref{thetaf}) this gives the value of $c$ needed,
\beq
c=\frac{\theta_f\omega_0}{[\pi^2+\theta_f^2]^{1/2}},
\eeq
whereas
\beqa
t_f&=&\frac{\pi}{\omega_1}=\frac{\pi}{\sqrt{\omega_0^2-c^2}}=\frac{\pi}{\omega_0}f,
\\
f&:=&\sqrt{1+\frac{\theta_f^2}{\pi^2}}.
\eeqa
As $c<\omega_0$ the effect of this bang-bang protocol is to expand the effective trap 
during the rotation time interval. $b$ increases first and then decreases during half an oscillation period
of the effective trap. This does not in general coincide with half oscillation period of the actual 
non-rotating trap $\pi/\omega_0$ because of the $f$ factor, but it is not too different for relevant values of $\theta_f$.  
In particular, for $\theta_f=\pi/2$, $f=1.118$.  The maximum of $b(t)$ at $t_f/2$ is precisely $f$. For example,  
for a frequency $\omega_0/(2\pi)=2$ MHz, this implies a final time $t_f=0.28$ $\mu$s.
%
% % % % % % % % % % % % % % % % % % % % % % % % % % % % % % % % % % % % % % % % %
%
\section{Optimal Control by Pontryagin's maximum Principle}
%
% \subsection{Optimization Problem with Fixed Boundary Conditions}
While the previous bang-bang method with just one time segment provides a simple guidance, we are also interested in
knowing the absolute time minimum that could in principle be achieved (even if the ``optimal'' protocol ends up being  
hardly realizable).    
Unlike ordinary expansions/compressions, the  shortest time protocol for bounded control 
is not of a bang-bang form.  To find it 
%We apply here Pontygrin's maximum principle \cite{Pon}. 
we first rescale the time with $\omega_0$ by setting $\sigma=\omega_0 t$ for $t\in [0,t_f]$. 
%Below, we shall show how the solutions obtained by this method are $\omega_0$ independent. 
Now we set the variables 
\begin{align}
	x_1(\sigma) & = b(t)=b\Big(\dfrac{\sigma}{\omega_0}\Big), \nonumber\\
	% \label{eq:x2}
	x_2(\sigma) & = \frac{1}{\omega_0}\dot{b}\Big(\dfrac{\sigma}{\omega_0}\Big), \\% \frac{db\Big(\dfrac{s}{\omega_0}\Big)}{ds}, \\
	% \label{eq:x3}
	x_3(\sigma) & = \int_0^{\sigma} u(\tau) d\tau, \nonumber
\end{align}
where $u(\sigma)=u(\omega_0 t)=\frac{1}{\omega_0}\dot{\theta}(t)$ with $\sigma\in [0,\omega_0 t_f]$. Then, we can write a control system describing the Ermakov equation \eqref{Erm} and the constraints in \eqref{cond1}, \eqref{cond2} and \eqref{cond3}, and formulate the time-optimal control (OC) problem for rotation of a quantum particle on a line as
\beqa
	\label{eqsys1}
	\min_u J &=& \int_0^T 1d\tau,
 \nonumber\\ % = T, \nonumber \\
	{\rm such\; that} \quad {x}_1' &=& x_2,
 \nonumber\\
	\quad {x}_2'&=&\frac{1}{x_1^3}+(u^2-1)x_1, 
\nonumber\\
	\quad {x}_3'&=&u, 
\eeqa
where $T = \omega_0 t_f$ and the prime is a  derivative with respect to $\sigma$, 
with the boundary conditions
\beqa
x_1(0) & =& 1, \quad x_1(T) = 1,
\nonumber\\
x_2(0) & =& 0, \quad x_2(T) = 0,
\nonumber\\
x_3(0) & =& 0, \quad x_3(T) = \theta_f.
\label{bcx}
\eeqa
Note that we assume that the boundary conditions for $u$ at $t=0$ and $t=t_f$ can be fulfilled by the use of 
a sudden switch. 
%
%% ========================================================
%\subsection{Derivation of the Time-Optimal Control} %by Pontryagin's Minimum Principle}
%
%
% --------------------------------
\subsubsection{Unbounded Control}
\begin{figure}[t]
\begin{center}
\includegraphics[width=8cm]{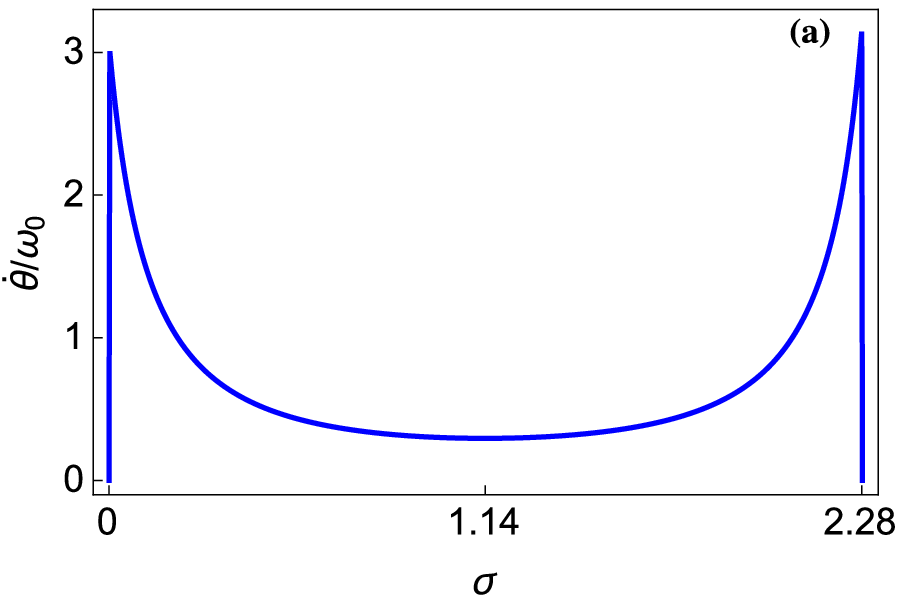}
\includegraphics[width=8cm]{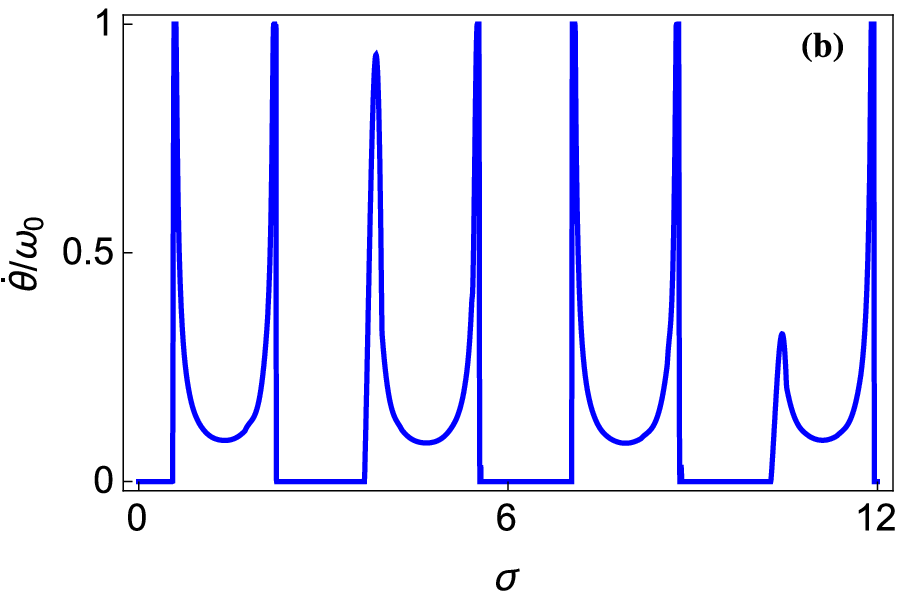}
\caption{\label{optimal}(Color online)
Time evolution $\dot{\theta}(t)$ for the optimal unbounded (a) and bounded (b) control. 
%$\omega_0/(2\pi)=2$ MHz, a $^9$Be atom, and 
The rotation angle is $\theta_f=\frac{\pi}{2}$.}
\end{center}
\end{figure}

We apply the Pontrygin's maximum principle \cite{Pon} to solve the time-optimal control problem \eqref{eqsys1}, where the Hamiltonian is given by
\beq
\label{eq:H}
H(t,x,u,\lambda) = \lambda_0+\lambda_1x_2+\lambda_2\Big[\dfrac{1}{x_1^3}+(u^2-1)x_1\Big]+\lambda_3u,
\eeq
in which $\lambda=(\lambda_0,\lambda_1,\lambda_2,\lambda_3)$ and $\lambda_0$ is either $0$ or $1$. The necessary condition $\frac{\partial H}{\partial u}=0$ gives
\beq
\label{eq:u*}
u^* = -\frac{\lambda_3}{2\lambda_2x_1},
\eeq
which minimizes the Hamiltonian and where the co-states $\lambda_1, \lambda_2, \lambda_3 : [0,T] \rightarrow \mathbb{R}$ satisfy ${\lambda}_i'=-\frac{\partial H}{\partial x_i}$, $i=1,2,3$, i.e.,
\beqa
\label{eqsys2}
	{\lambda}_1' & =& \Big[\dfrac{3}{x_1^4}-(u^2-1)\Big] \lambda_2,
 \nonumber\\ 
	{\lambda}_2' & =& - \lambda_1, 
\\ %\quad \lambda_2(T) = b,\\
	{\lambda}_3' & =& 0. \nonumber
\eeqa
Solutions are found by solving the equation system composed by Eqs. \eqref{eqsys1}, \eqref{eq:u*}, and \eqref{eqsys2}
with the boundary  conditions at $\sigma=0$ in Eq. (\ref{bcx}).  We have the freedom of choosing different initial values for the $\lambda_i(0)$ to satisfy the boundary conditions at $T$ in Eq. (\ref{bcx}). We apply a shooting method and numerically minimize $[x_1(T)-1]^2+x_2(T)^2+[x_3(T)-\theta_f]^2$ for these parameters using MATLAB's `fminsearch' function with $\theta_f=\pi/2=1.5708$. The best results obtained are for $T=2.2825$, which, for the external trap frequency $\omega_0/(2\pi)=2$ MHz used in other examples, implies a final time $t_f=0.18$ $\mu$s. The solution found is not exact, $(x_1(T), x_2(T), x_3(T))=(1.0765, 0.0842, 1.5650)$, which might be 
an indication that the system is not controllable. Figure \ref{optimal} (a) shows the time evolution of $u$ for this case following Eq. \eqref{eq:u*} but forcing it to be 0 in the boundary times.

% % % % % % % % % % % % % % % % % % % % % % % % % % % % % % % % % % % % % % % % %
% % % % % % % % % % % % % % % % % % % % % % % % % % % % % % % % % % % % % % % %
% % % % % % % % % % % % % % % % % % % % % % % % % % % % % % % % % % % % % % % % % % %
% % % % % % % % % % % % % % % % % % % % % % % % % % % % % % % % % % % % % % % % % % %
%
% ------------------------------------
\subsubsection{Bounded Control}
Now,  consider a bounded control with $u(\sigma)\in [0,1]$ for all $\sigma\in [0,T]$. Because the Hamiltonian  \eqref{eq:H} is quadratic in $u$, the optimal control that minimizes $H$ is of the form
\beq
	\label{eq:u_bdd}
	u^*_b = \min\bigg\{ \max \left\{ -\dfrac{\lambda_3}{2\lambda_2x_1},0\right\} , 1\bigg\}.
\eeq
The bounded time-optimal control and the resulting optimal trajectory 
are illustrated in Figure \ref{optimal} (b). The minimum (dimensionless) time that completes the desired rotation is $T=11.9984$ and the calculated final state following the optimal control is $(x_1(T), x_2(T), x_3(T))=(1.0083, 0.0382, 1.5708)$. For $\omega_0/(2\pi)=2$ MHz, the minimal time is  $0.95$ $\mu$s.  Since $u(\sigma)\in [0,1]$, $\forall\ \sigma\in [0,T]$, from \eqref{eqsys1} we see that $\dot{\theta}>0$, and hence the rotation is always forward. In this case, $x_3$ reaches the desired $\theta_f=\pi/2$ at $\sigma=11.9028$, and the control is turned off. Then, the states $x_1$ and $x_2$ are oscillating to reach the desired terminal state $(1,0)$. Figure \ref{optimal} (b) shows the time evolution of $u$ for this solution. 
%
%
%
%
%
%
%
%
% % % % % % % % % % % % % % % % % % % % % % % % % % % % % % % % % % % % % % % % % % %
% % % % % % % % % % % % % % % % % % % % % % % % % % % % % % % % % % % % % % % % % % %
%
%
%
\section{Smooth inverse engineering}
\begin{table}[b]
\centering
\begin{tabular}{ l | c c c c }
\hline \\
   & bang-bang &  OC(unbounded) & OC(bounded) & inverse engineering \\ \hline
$t_f$ ($\mu$s) & 0.28 & 0.18 & 0.95 & 0.23 \\ \hline
\end{tabular}
\caption{Minimal rotation times for the different methods. Trap frequency $\omega_0/(2\pi)=2$ MHz. In bounded OC,
$0\leq \dot{\theta}\leq\omega_0$.} 
\label{table} 
\end{table}%
An alternative inversion route that provides smooth solutions is depicted in
the following scheme
\begin{displaymath}
    \xymatrix{
        \theta \ar[r]& \dot{\theta} \ar[r]&\omega \ar[r]&E[b(t_f),\dot{b}(t_f)]  \ar@/^2pc/@{.>}[lll]_{minimize\; E}
        }
\end{displaymath}      
First, $\theta(t)$ is designed to satisfy Eq. (\ref{cond1}) and Eq. (\ref{cond2}) with some free parameters. 
The corresponding $\dot{\theta}$ and final energy are calculated, and the parameters are changed until 
the minimum energy (and excitation) is found.  

% % % % % % % % % % % % % % % % % % % % % % % % % % % % % % % % % % % % % % % %
\begin{figure}[t]
\begin{center}
\includegraphics[width=8cm]{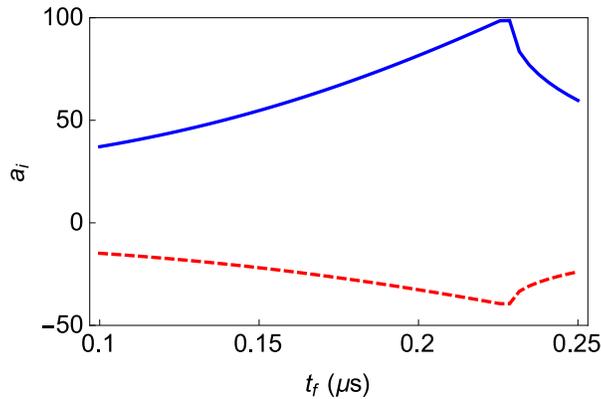}
\caption{\label{rotparams}(Color online)
Values of the optimizing parameters $a_4$ (thick blue line) and $a_5$ (dashed red line) for different rotation times $t_f$. 
%In the simulations we used a trap of frequency 
%$\omega_0/(2\pi)=2$ MHz, a $^9$Be atom, and a rotation angle 
The trap frequency is $\omega_0/(2\pi)=2$ MHz, and 
the final angle $\theta_f=\frac{\pi}{2}$.}
\end{center}
\end{figure}
% % % % % % % % % % % % % % % % % % % % % % % % % % % % % % % % % % % % % % % % % % %
 
% % % % % % % % % % % % % % % % % % % % % % % % % % % % % % % % % % % % % % % %
\begin{figure}[t]
\begin{center}
\includegraphics[width=8cm]{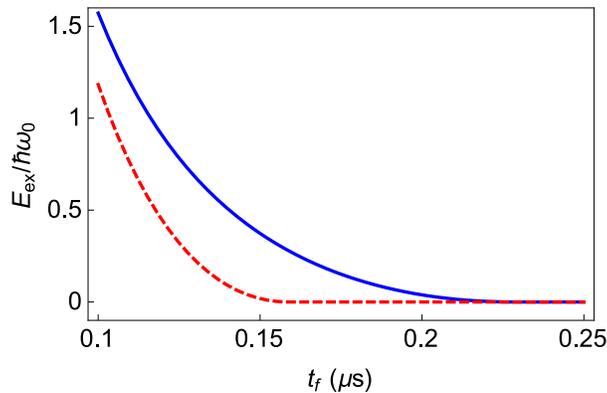}
\caption{\label{excitation}(Color online)
Final excitation energy vs final time for the optimized protocol without (solid blue line) and with final squeezing 
($\gamma^2=3$, dashed red line). The trap  frequency is 
$\omega_0/(2\pi)=2$ MHz, 
%a $^9$Be atom, a 
and  the final rotation angle $\theta_f=\frac{\pi}{2}$. The initial state is the ground state of the trap.}
\end{center}
\end{figure}
%
% % % % % % % % % % % % % % % % % % % % % % % % % % % % % % % % % % % % % % % % %
% % % % % % % % % % % % % % % % % % % % % % % % % % % % % % % % % % % % % % % %
\begin{figure}[t]
\begin{center}
\includegraphics[width=8cm]{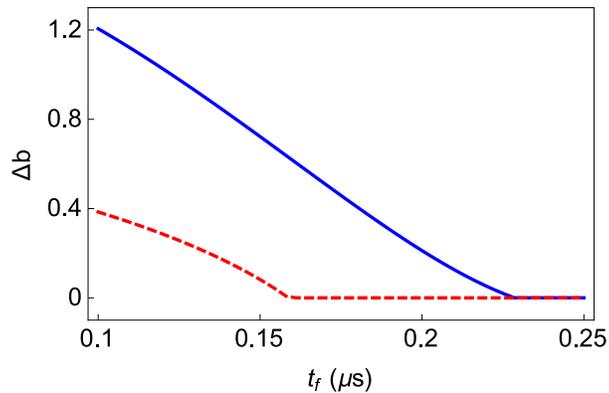}
\caption{\label{difb}(Color online)
Difference between ideal and actual value of $b$ at the end of the rotation vs final time for the optimized inverse-engineered protocol for rotations without (solid blue line) and with final squeezing ($\gamma^2=3$, dashed red line). 
The trap  frequency
is $\omega_0/(2\pi)=2$ MHz, 
%a $^9$Be atom, 
and the final rotation angle $\theta_f=\frac{\pi}{2}$.}
\end{center}
\end{figure}
% % % % % % % % % % % % % % % % % % % % % % % % % % % % % % % % % % % % % % % % % % %
% % % % % % % % % % % % % % % % % % % % % % % % % % % % % % % % % % % % % % % % % % %
%

% % % % % % % % % % % % % % % % % % % % % % % % % % % % % % % % % % % % % % % % % % %% % % % % % % % % % % % % % % % % % % % % % % % % % % % % % % % % % % % % % % %
\begin{figure}[t]
\begin{center}
\includegraphics[width=8cm]{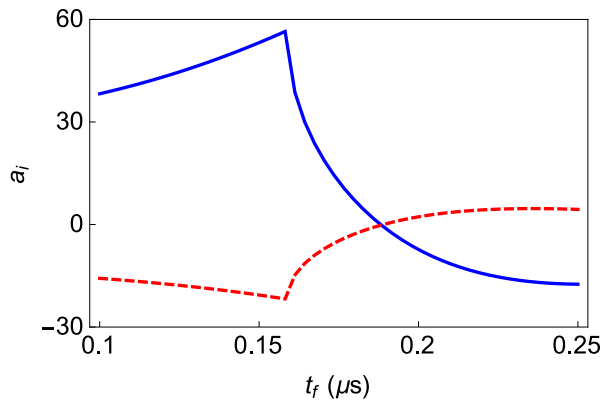}
\caption{\label{squparams}(Color online)
Values of the optimizing parameters $a_4$ (thick blue line) and $a_5$ (dashed red line) for different rotation times to generate a squeezed vaccum state with $\gamma^2=3$. 
The trap  frequency is 
$\omega_0/(2\pi)=2$ MHz,   
%a $^9$Be atom, 
and the rotation angle $\theta_f=\frac{\pi}{2}$.}
\end{center}
\end{figure}
% =================================================

A convenient choice for $\theta$ is a fifth order polynomial ansatz $\theta~=~\sum_{n=0}^5 a_n t^n/t_f^n$. 
%, where $s=t/t_f$. 
In order to satisfy the boundary conditions in Eqs. (\ref{cond1}) and (\ref{cond2}) we need to fix parameters $a_{0-3}=(0,0,a_4+2a_5+3\theta_f,-2a_4-3a_5-2\theta_f)$. The other two parameters, $a_4, a_5$, are left free in order to satisfy the remaining two boundary conditions in Eq. (\ref{finalcb}) and suppress the final excitation energy. In 
%principle, one could expect to exactly satisfy these conditions for arbitrary final times by solving $b$ in the differential Eq. (\ref{differential}), and then fixing $a_4, a_5$ to satisfy Eq. (\ref{finalcb}). However, it is not a simple differential equation to solve analytically. Thus, we restricted ourselves to a numerical solution. The procedure was as follows.
practice we solve numerically Eq. (\ref{differential}) to find the final energy  (\ref{energy}) for each pair $a_4,a_5$, and use 
MATLAB's `fminsearch' function to find the values of the free parameters that minimize the final excitation energy.
%a shooting 
% We first tried some ansatz values for the free parameters, that define $\theta$ and $\dot{\theta}$. Then, we numerically solved for $b$ and obtained the final energy from . We repeated the calculation within a shooting program by  

In Fig. \ref{rotparams} the values of the free parameters that result from this process are given, and in Fig. \ref{excitation} 
we depict the corresponding excess energy with respect to the ideal target state (as in previous examples, $\omega_0/(2\pi)=2$ MHz). Vanishing residual excitations are found for 
times shorter than half an oscillation period up to a time $t_f\sim 0.23$ $\mu$s, not much larger than the unbounded-optimal-control minimum  of $0.18$ $\mu$s. Fig. \ref{difb} depicts the difference between the ideal value of $b(t_f)$ and the actual value, and makes evident the sharp change 
that marks the shortest time for which a solution exists. Since we have limited the possible solutions by imposing a functional form of the function $\theta (t)$, this time is larger than the one found via OC. Note also that the shortest final time is  slightly better than the one provided by the simple bang-bang protocol. Table I summarizes the results. 
\section{Wave packet squeezing}
Consider now a trap rotation with constant trap frequency $\omega_0$ satisfying the conditions  (\ref{cond1}-\ref{cond3}), 
and $b$ satisfying 
\beqa
b(0)&=&1,\;\;\,\dot{b}(0)=0,
\nonumber
\\
b(t_f)&=&\gamma,\; \dot{b}(t_f)=0.  
\label{bsq}
\eeqa
Unlike the previous sections, $b$ ends in a value $\gamma$ different from 1. 

According to Eq. (\ref{psin}), each initial state $\phi_n(0)$ will evolve into $e^{-i(n+1/2)\omega_0g}\phi_{n,sq}$ at $t_f$,
where  $g=g(t_f)=\int_0^{t_f} dt'/b^2(t')$, and $\phi_{n,sq}$ is the normalized eigenstate for the trap with angular frequency
$\omega_{sq}=\omega_0/\gamma^2$.  
(This is a virtual trap, let us recall that the actual trap has angular frequency $\omega_0$.)   

A coherent state at time $t=0$, 
\beq
|\alpha\ra=e^{-|\alpha|^2/2}\sum_{n=0}^\infty \frac{\alpha^n}{\sqrt{n!}}|\phi_n(0)\ra,
\eeq
will thus evolve into 
\beq
|\psi(t_f)\ra=e^{-i\omega_0g/2}e^{-|\tilde{\alpha}|^2/2}\sum_{n=0}^\infty \frac{\tilde{\alpha}^n}{\sqrt{n!}}|\phi_{n,sq}\ra,
\eeq
where $\tilde{\alpha}=\alpha e^{-i\omega_0g}$. This is a coherent state for the virtual frequency $\omega_{sq}$
and therefore a minimum-uncertainty-product state. 
However, since the actual trap has frequency $\omega_0$, it is also a squeezed coherent state with respect to the actual trap, $|[r,\tilde{\alpha}]\ra$, see
\cite{Wolf}, where $r=-\ln \gamma$, up to a global 
phase factor. The final and initial coordinate and momentum widths are related by $\Delta_{s,t_f}=\gamma \Delta_{s,0}$, $\Delta_{p,t_f}=\Delta_{p,0}/\gamma$.   
%and therefore unstable: from $t_f$ on it will breath back and forth   
We may rewrite the state at time $t_f$ in terms of the squeezing and displacement operators as  
\beq
|\psi(t_f)\ra=e^{-i\omega_0g/2}S(r)|\tilde{\alpha}\ra=e^{-i\omega_0g/2}S(r)D(\tilde{\alpha})|0\ra=e^{-i\omega_0g/2}|[r,\tilde{\alpha}]\ra,
\eeq
where $S(r)=e^{\frac{r}{2}(a^2-{a^\dagger}^2)}$, $a$ and $a^\dagger$ are annihilation and creator operators for the $\omega_0$-harmonic trap, 
and $D(z)=e^{za^\dagger-z^*a}$ is the displacement operator.    
Note that  the phase at $t_f$, $\arg(\tilde{\alpha})$, is controllable by means of the $g$-function that depends on the process history, 
whereas the squeezing parameter $1/\gamma$ is controlled by the imposed boundary condition.  If necessary, a  controlled tilt of the squeezed state in phase space is easy to achieve by letting it evolve, after its formation at $t_f$, in the fixed, non-rotating  trap.      
 
As a simple example let us consider the generation of squeezed vacuum states starting from the ground state of the initial trap, 
so that $\alpha=0$.   
To design the squeezing process  
we may follow a similar procedure as in the previous section, but minimizing  the cost function
\beqa
F&=&
%\frac{2n+1}{4 \omega(0)}\left(
\dot{\tilde{b}}(t_f)^2+\omega ^2(t_f)\tilde{b}(t_f)^2+\frac{\omega (0)^2}{\tilde{b}(t_f)^2},
%\right),
\nonumber\\
\tilde{b}&=&b-\gamma +1,
\eeqa
%
%{\ES EXPLAIN WHY THESE FORMS of F and $\tilde{b}$}
which is minimal for 
%be a stationary state for a trap with a frequency $\omega_{sq}=\omega_0/\gamma ^2$  
$\tilde{b}(t_f)=1$ and $\dot{\tilde{b}}(t_f)=0$, so that $b(t_f)=\gamma$  and $\dot{b}(t_f)=0$.
%This corresponds to a stationary state for a trap with a frequency $\omega_{sq}=\omega_0/\gamma ^2$.  
%This cost function looks like the expresion for the final energy obtained from Eq. \eqref{energy}, but with this new function $\tilde{b}$ instead of $b$. Thus, the function will be minimized when $b=\gamma$, which is the corresponding to the ground state in a trap of frequency $\omega_{sq}$. Another way of understanding this cost function, is by interpreting it as the divergence final energy between the state corresponding to the frequency $\omega_{sq}$ and the obtained state. The smaller it is, the closer we will be to that state.
%At $t_f$ the trap has angular frequency $\omega_0$ so the squeezed state may be stabilized by a sudden  jump to $\omega_{sq}$. 

Since, due to the centrifugal force during the rotation, the wave packet tends to spread first, the squeezed states with $\gamma>1$ may be achieved in shorter times 
than the ones needed without squeezing in the previous section.   
Figure \ref{squparams} depicts the free parameters that optimize a rotation with a final squeezed state for the same parameters in the previous subsection, but $\gamma^2=3$, 
and Fig. \ref{excitation} the excess energy with respect to the target state. The excitation in a process with a final moderate squeezing is  smaller than for the simple rotation without squeezing. Fig. \ref{difb} depicts the difference between the target value of the function $b$ (proportional to the width of the wavepacket) and its actual value at final time for rotations without and with squeezing. Again, the minimizations change suddenly to a different solution
that cannot satisfy the conditions at a critical time, see also Figs. \ref{rotparams} and \ref{squparams}. 
%At these time we see a singularity, produced by the sudden increase in the divergence between the target and the actual values of $b$.

%
%
%
\section{Discussion}
We have worked out different schemes to perform fast rotations of a one-dimensional trap
without any final excitation of the confined particle, which we have considered to be an ion 
throughout but could be a neutral particle as well by setting the proper trapping interaction. 
Apart from excitation-free rotations it is also possible to generate  squeezed states in a controllable way. 
For an arbitrary trap,  the fast processes could in principle be performed 
in an arbitrarily short time if an auxiliary harmonic potential with time dependent  frequency could be implemented. 
In a simpler setting, where only the rotation speed may be controlled, the rotation time cannot be arbitrarily short, as demonstrated  by inverse engineering or bang-bang approaches, and confirmed by optimal-control theory. 
Bang-bang and optimal control protocols provide useful information and time bounds but  are difficult to implement experimentally due to the sudden kicks in the angular velocity of the trap. Smooth protocols designed by invariant-based inverse engineering 
have also been worked out. They achieve negligible excitations for times close to the minimum times given by optimal control theory. 

The analysis may be generalized for a two-dimensional trap but it becomes considerably
more involved \cite{Masuda} and will be considered separately.  
The 1D approximation used here will be valid 
for total energies well below the transversal confinement energy $E_{\perp}=\hbar\omega_{\perp}$. 
For the shortest final times considered  in our simulations, excitation energies are never larger than $2\hbar\omega_0$
so that $\omega_{\perp}\gg \omega_0$ would be enough for their validity. 

Rotations are elementary manipulations which together with transport, splitting, and expansions, may 
help to build a scalable quantum information architecture.  In particular, they provide a mechanism for connecting 
sites by changing transport directions in 2D networks.  
Rotations have been demonstrated experimentally for trapped ions \cite{Splatt} and improving the capability to control the parameters involved is feasible with state-of-the-art trapped-ion technology.    
To extend the present analysis to ion chains  \cite{Splatt}, an approach similar to that in \cite{Palmerotrans2,Palmeroexp,Palmerosep} could be applied, working out the dynamical modes of the system and taking into account the dipole-dipole interaction due to the rotation of the charged particles. 
The present  results set a first step towards accurately controlling rotating ion chains which would allow for
fast reordering.    
 
%Simulations show that, for three of the methods, namely the Bang-Bang, the unbounded optimal control and the inverse engineering we are able to perform a $\pi/2$ rotation in as fast as half an oscillation period just by wisely designing the time dependent rotation angle. Another possibility, , but it requieres the manipulation of two control parameters (angle and frequency of the trap) instead of the single control parameter (angle) as in the other methods. 

%The ability to design fast rotations is important in quantum mechanics in general, since it is a basic operation we can perform in many different situations. In particular, in trapped ions, it is interesting for a hypothetical scalable quantum processor as the NIST-built \emph{racetrack} \cite{Amini}, where one should send the ions through Y junctions, and presumably rotate them. 

%A possible extension of this work is to study a rotation within a 2-dimensional trap as done in Ref. \cite{Masuda}. Another possibility is to study rotations of ion chains as in Ref. \cite{Splatt}, wich would allow to reorder this ion chain. %without the need of actually breaking the chain.

%
%
%
%
\section*{Acknowledgements}
We thank Kazutaka Takahashi, Joseba Alonso, Uli Poschinger, and Christian Schmiegelow for useful discussions.  
This work was supported by 
the Basque Country Government (Grant No. IT472-10), 
Ministerio de Econom\' \i a y Competitividad (Grant No. FIS2012-36673-C03-01), 
the program UFI 11/55 of the Basque Country University 
and the US National Science Foundation under the awards CMMI-1301148 and CMMI-1462796.
M.P. acknowledges a fellowship by UPV/EHU.
\appendix
\section{Wave functions\label{app}}
The time-dependent wave functions evolving with the Hamiltonian (\ref{Ham})
take the form \cite{LR,Chen,Torrontegui}
\beq
\la s|\psi(t)\ra=\sum_n c_n e^{i\alpha_n(t)} \la s|\phi_n(t)\ra
\eeq
where the $c_n$ are constant,  
\beqa
\alpha_n(t)&=&-\frac{1}{\hbar}\int_0^t dt' \frac{(n+1/2)\hbar\omega_0}{b^2}=-\omega_0(n+1/2)\int_0^t dt' \frac{1}{b^2},
%\eeq
%
\\
%\beq
\label{psin}
\la s|\phi_n(t)\ra&=&e^{\frac{im}{\hbar}\dot{b}q^2/(2b)}\frac{1}{b^{1/2}}\Phi_n(s/b),
\eeqa
and $\Phi_n(x)$ is the Hermite polynomial solution of the harmonic oscillator with angular frequency $\omega_0$
and energy eigenvalue $(n+1/2)\hbar \omega_0$, 
$\Phi_n(x)=\frac{1}{\sqrt{2^n n!}}(\frac{m\omega_0}{\pi\hbar})^{1/4} e^{\frac{-m\omega_0 x^2}{2\hbar}} H_n(\sqrt{\frac{m\omega_0}{\hbar}} x)$. Note that $\frac{1}{b^{1/2}}\Phi_n(s/b)$ is just a scaled state which corresponds to the $n$-th eigenstate of a trap 
with angular frequency $\omega_0/b^2$.

\end{document}